\documentclass[12pt]{article}
\usepackage[margin=0.75 in]{geometry}
\usepackage{amsmath,amssymb}
\usepackage{graphicx}
\usepackage{float}
\usepackage{caption}
\usepackage{hyperref}
\usepackage{setspace}
\def\dd{{\rm d}}\def\ee{{\rm e}}\def\_#1{^{}_{#1}}
\def\beq{\begin{equation}}\def\eeq{\end{equation}}
\def\bea{\begin{eqnarray}}\def\eea{\end{eqnarray}}
\begin{document}
\onehalfspacing
\title{\sc Path Length Distribution\\ in Two-Dimensional Causal Sets}
\author{M. Aghili\thanks{maghili@go.olemiss.edu}, L. Bombelli\thanks{bombelli@olemiss.edu},
B.B. Pilgrim\thanks{bbpilgri@go.olemiss.edu}}
\date{May 18, 2018}
\maketitle
\abstract{\noindent We study the distribution of maximal-chain lengths between two elements of a causal set and its relationship with the embeddability of the causal set in a region of flat spacetime. We start with causal sets obtained from uniformly distributed points in Minkowski space. After some general considerations we focus on the 2-dimensional case and derive a recursion relation for the expected number of maximal chains $n_k$ as a function of their length $k$ and the total number of points $N$ between the maximal and minimal elements. By studying these theoretical distributions as well as ones generated from simulated sprinklings in Minkowski space we identify two features, the most probable path length or peak of the distribution $k_0$ and its width $\Delta$, which can be used both to provide a measure of the embeddability of the causal set as a uniform distribution of points in Minkowski space and to determine its dimensionality, if the causal set is manifoldlike in that sense. We end with a few simple examples of $n_k$ distributions for non-manifoldlike causal sets.

\section{\sc Introduction}
In the causal set approach to quantum gravity, the Lorentzian manifold used in general relativity and other theories of gravity to represent the spacetime geometry is simply the large-scale view of a locally finite partially ordered set, a causal set \cite{sorbom}. Part of the motivation for this approach is the observation \cite{myrheim,bomb02} that if one chooses a sufficiently dense set of uniformly distributed random points in a spacetime manifold, one can recover the spacetime geometry on scales larger than the one determined by the point density simply by using the causal ordering of the points. In this view then, spacetime is a purely combinatorial structure, a collection $\mathcal{C}$ of events with a relation $p\prec q$ which is reflexive, transitive, symmetric, and makes $\mathcal{C}$ locally finite in the sense that for any $p$, $q\in \mathcal{C}$ the interval or Alexandrov set $I(p,q):= \{r\mid p\prec r\prec q\}$ has only a finite number of elements; for the purposes of this paper, without loss of generality from now on we will assume that $\mathcal{C}$ is actually of finite cardinality and corresponds to a spacetime region of finite volume.

Given a Lorentzian manifold without closed causal curves, any choice of a finite number $N$ of points in it, with the partial order induced by causality, will produce a causal set; if the points are randomly distributed with uniform density, the causal set samples the continuum geometry uniformly and can be seen as a discretization of it at the volume scale given by the inverse density provided there are no length scales in the manifold of the order or smaller than the average distance between the points. On the other hand, given a large number $N$, the vast majority of the causal sets that can be constructed out of $N$ elements cannot arise as discretizations of Lorentzian manifolds. One of the most general questions causal set theory must address then is what makes it possible, in this approach, for the causal sets arising out of the dynamics to be such that they are seen as Lorentzian manifolds at large scales. Some preliminary steps towards answering this question consist in showing that if a large causal set is manifoldlike then the manifold is approximately unique in an appropriate sense \cite{bomb02,bomb03}, establishing criteria for manifoldlikeness of causal sets, possibly involving their embeddability in Lorentzian manifolds \cite{henson, major, glasersurya, eichhorn, yazdi}, and identifying procedures for using the structure of a manifoldlike causal set to determine properties of the corresponding manifold, such as its dimensionality, topology or curvature \cite{meyer,bright,surya,bendowker}.

In this paper, we use the distribution of maximal-chain lengths between pairs of points in a causal set to find a criterion for manifoldlikeness. As shown in Section 3, the path distribution has an approximately Gaussian shape, with a maximum and a width that can be used as two parameters characterizing the structure of that interval in the causal set.

\section{\sc The Path Length Distribution}
In this section we will derive an expression for the distribution of maximal-chain lengths in a causal set sprinkled uniformly at random in an Alexandrov set $I(p,q)$ of 2-dimensional Minkowski spacetime, and show the results of some numerical simulations. This distribution provides a good opportunity for statistical analysis of properties of the causal set. We will refer to an element of such a sprinkling that is contained in some smaller region within $I(p,q)$ simply as a point in that region, and we will call a region empty if it does not contain any of the randomly distributed points. A maximal chain of length $k$ between points $x\_1$ and $x\_{k+1}$ is defined by $k+1$ related points $x\_1 \prec x\_2 \prec ... \prec x\_k \prec x\_{k+1}$ such that for each $i$ the Alexandrov set $I\_{i,i+1} = I(x\_i,x\_{i+1})$ is empty; in the following, a maximal chain will be called a {\em path}.

The number of $k$-paths between two points $p = x\_1$ and $q = x\_{k+1}$ in a causal set uniformly embedded in Minkowski space is a random variable, whose mean value can be evaluated analytically by picking $k-1$ possible locations for the other $x\_i$, calculating the probability that one causal set point is found in each infinitesimal neighborhood $\dd x\_i$ and each interval $I\_{i,i+1}$ is empty, and integrating over all the $x\_i$.

When $N$ points are sprinkled uniformly in a volume $V_0$, the probability of finding exactly $k$ of them within a region of volume $V$ inside it is given by the binomial distribution,
\beq
P_k = {N\choose k}\left(\frac{V}{V_0}\right)^{\!k}\left(1-\frac{V}{V_0}\right)^{\!N-k}.
\eeq
In the small $V/V_0$ limit, as in the case of an infinitesimal $\dd x\_i$, this probability can be approximated by a Poisson distribution of density $\rho = N/V_0$. Thus, for each of the $\dd^d x_i$ the probability that it contains exactly one point can be written as $\rho\,\dd^d x\_i\,\ee^{-\rho\,\dd^d x\_i} \approx \rho\,\dd^d x_i$, and all of those probabilities can be considered to be independent as long as the number of links in the chain is much smaller than the total number of points, $k \ll N$. The intervals $I\_{i,i+1}$ however may not be small, and for those we will use the binomial distribution. In particular, the probability that a region of volume $V$ is empty is given by $P\_0 = (1-V/V_0)^N$, and for the union $I\_{1,2}\cup \cdots \cup I\_{k,k+1}$ of all Alexandrov sets between pairs of points that probability can be written as
\beq
P_0=\left(1-\frac{\sum_{i=1}^k V\_{i,i+1}}{V_0}\right)^{\!N}.
\eeq
Putting these together we then get, following the same approach as in Ref.\ \cite{meyer},
\beq
P(x_2,\ldots,x_k)\,\dd^d x_2\,\cdots\,\dd^d x_k
=\rho^{k-1}\,\dd^dx_2\ \cdots \dd^dx_k\ \left(1-\frac{\sum_i^k V_{i,i+1}}{V_0}\right)^{\!N}
 + \hbox{higher-order terms}\;. \label{ppath}
\eeq
We now identify the probability in Eq.\ \ref{ppath} with the mean number of paths through those locations, which integrated over all $x_i$ gives the mean number of $k$-paths between $p$ and $q$,
\beq
\langle n_k\rangle = \rho^{k-1}\int_{I_1}\dd^dx_2\cdots\int_{I_{k-1}}\dd^dx_k~\left(1-\frac{\sum_i ^k V_{i,i+1}}{V_0}\right)^{\!N},
\eeq
where $I_i = I(x\_i,q)$ is the Alexandrov set between $x\_i$ and the maximal element $q$ in the manifold; for simplicity, from now on we will drop the angle brackets, $\langle n_k\rangle \mapsto n_k$. Using the binomial expansion we can write
\bea
&& n_k = \rho^{k-1}
\sum_{n=0}^N {N \choose n}\left({-\frac{1}{V_0}}\right)^{\!n} \times\nonumber\\
&&\kern50pt\sum_{i_1=0}^n {n\choose i_1} \cdots \sum_{i_{k-1}=0}^{i_{k-2}} {i_{k-2} \choose i_{k-1}}\int_{I_1}\dd^dx_2\cdots \int_{I_{k-1}}\dd^dx_k\, (V_{12})^{n-i_1}\cdots (V_{k,k+1})^{i_{k-1}}\;.\qquad \label{nk}
\eea

In two dimensions the volume of an Alexandrov set can be easily calculated using the null coordinates
\beq
u = (t+x)/\sqrt{2}\;, \qquad v = (t-x)/\sqrt{2}\;,
\eeq
in terms of which
\beq
V_{i,i-1} = (u_i-u_{i-1})(v_i-v_{i-1})\;. \label{null}
\eeq
With these expressions for the volumes, in the $d = 2$ case the integrals in Eq.\ \ref{nk} give
\begin{equation}
n_k = N^{k-1}\sum_{i=0}^N{N\choose i}(-1)^i\frac{\Gamma(i+1)}{\Gamma(i+k)^2}\,f_{i,k} \label{nk2}
\end{equation}
where we used $N = \rho V_0$ and
\begin{eqnarray}
&&f\_{i,k}=\sum_{i_2=0}^i\Gamma(1+i-i_2)\times\nonumber\\
&&\kern30pt\underbrace{\sum_{i_3=0}^{i_2}\Gamma(1+i_2-i_3)\cdots\underbrace{\sum_{i_{k-1}=0}^{i_{k-2}}\Gamma(1+i_{k-2}-i_{k-1})\underbrace{\sum_{i_k=0}^{i_{k-1}}\Gamma(1+i_{k-1}-i_k)\,\Gamma(i_k+1)}_{f\_{i_{k-1},2}}}_{f\_{i_{k-2},3}}}_{f\_{i_2,k-1}}\;.\qquad \label{fik1}
\end{eqnarray}
As suggested by the underbraces, this definition implies the recursion relation
\begin{equation}
f\_{i,k}=\sum_{j=0}^i\Gamma(1+i-j)f\_{j,k-1}\;, \label{fik2}
\end{equation}
with $f\_{i,1}:= \Gamma(i+1).$ Using Eqs.\ \ref{nk2}--\ref{fik2}, $n\_k$ may now be calculated for any $N$ and $k$.

\section{\sc Results of Simulations}
We wish to compare the results of the analytical distribution with actual manifoldlike causal sets obtained from numerical simulations of random sets of points sprinkled with uniform density in the Alexandrov set defined by two timelike related points in Minkowski space.  From Fig.\ \ref{sprink} it's easy to see that the theory matches well with the average of the simulations, though it is worth pointing out that as the large error bars suggest, individual sprinklings can deviate significantly from the theory. This problem can be somewhat though not completely mitigated by considering only the peak and width\footnote{By width we mean full width at half maximum.} of the distribution rather than its entirety. The shape of these distributions is nearly Gaussian, allowing us to characterize each curve with just these two numbers. As one can see in Fig.\ \ref{param} the relative errors in both the peak position and the width decrease with $N$, implying that the larger errors in Fig.\ \ref{sprink} are primarily due to fluctuations in the total number of paths rather than the shape of the curve considered as a probability distribution; however, there is still enough error in the peak and width to cause us some concern. As a result, any application based on this distribution should account for statistical fluctuations in evaluating a single causal set. For instance, if one wishes to use this distribution for its stated purpose of determining manifoldlikeness of causal sets, the failure of a particular causal set to exactly match either the full analytical distribution or its peak and width should \textit{not} be taken as a sign that the causal set is not manifoldlike; rather, a fairly large range around the analytical distribution should be used, and causal sets which fall into this range should be considered candidates for manifoldlike causal sets. We will discuss this further in the next section.

\begin{figure}
\begin{minipage}{0.45 \textwidth}
\centering
\includegraphics[width = \textwidth]{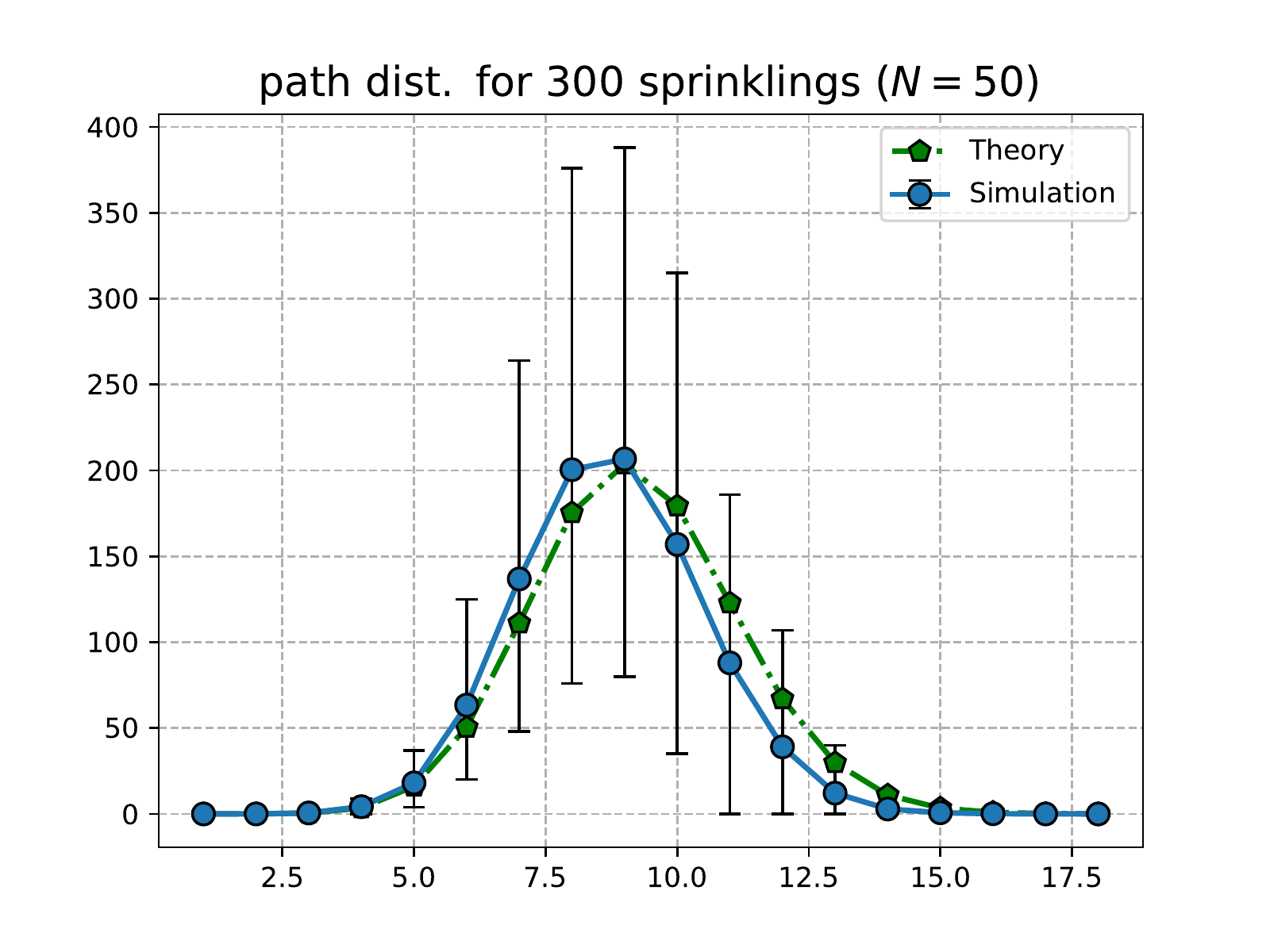}
\end{minipage}
\begin{minipage}{0.45 \textwidth}
\centering
\includegraphics[width = \textwidth]{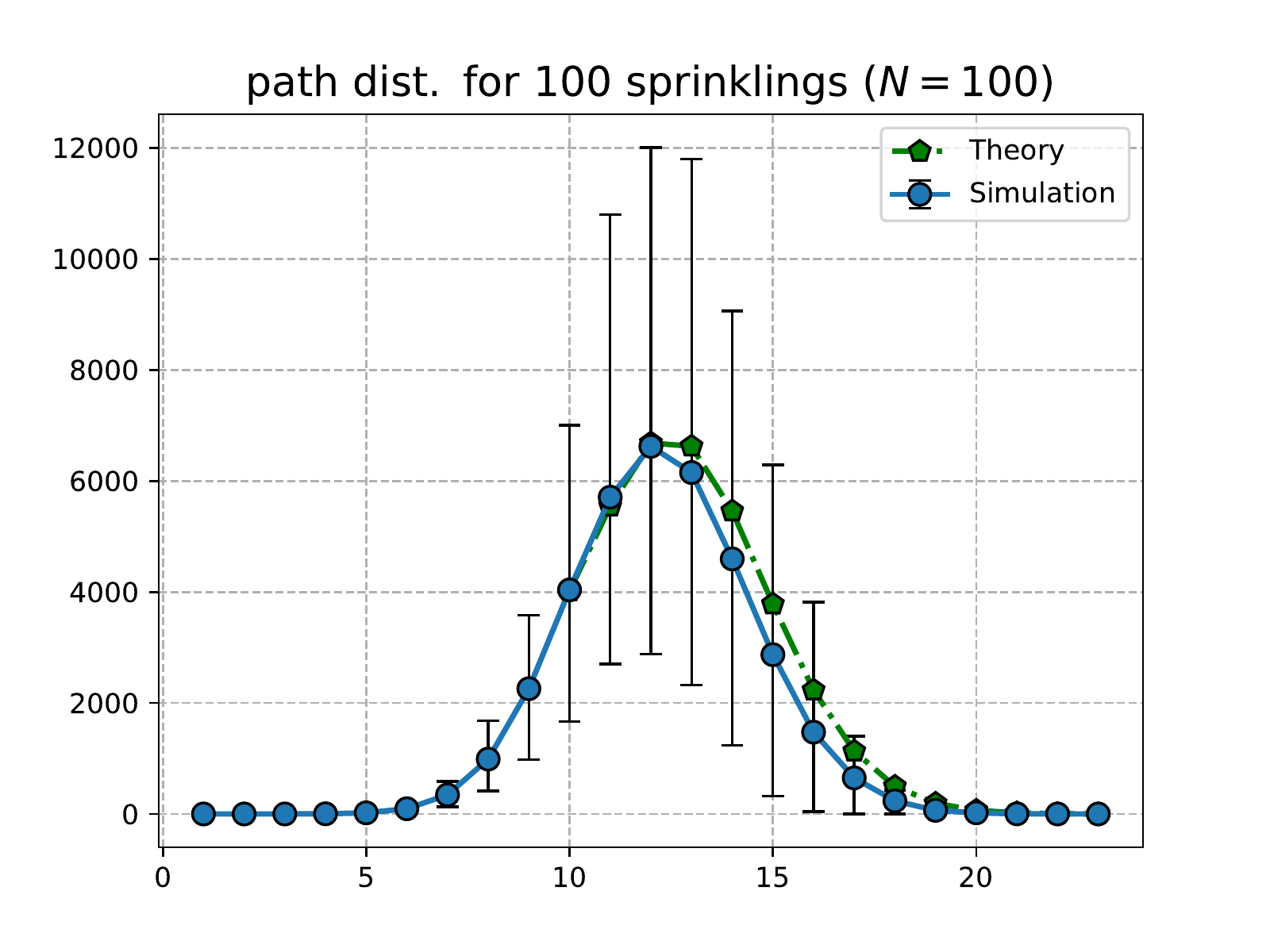}
\end{minipage}
\caption{A comparison of the average of many sprinklings of 50 and 100 points to their corresponding analytical distributions, on the left and right respectively. Due to numerical error, the theory curve starts at 9. \label{sprink}}
\end{figure}
\begin{figure}
\centering
\includegraphics[width = 3.6 in]{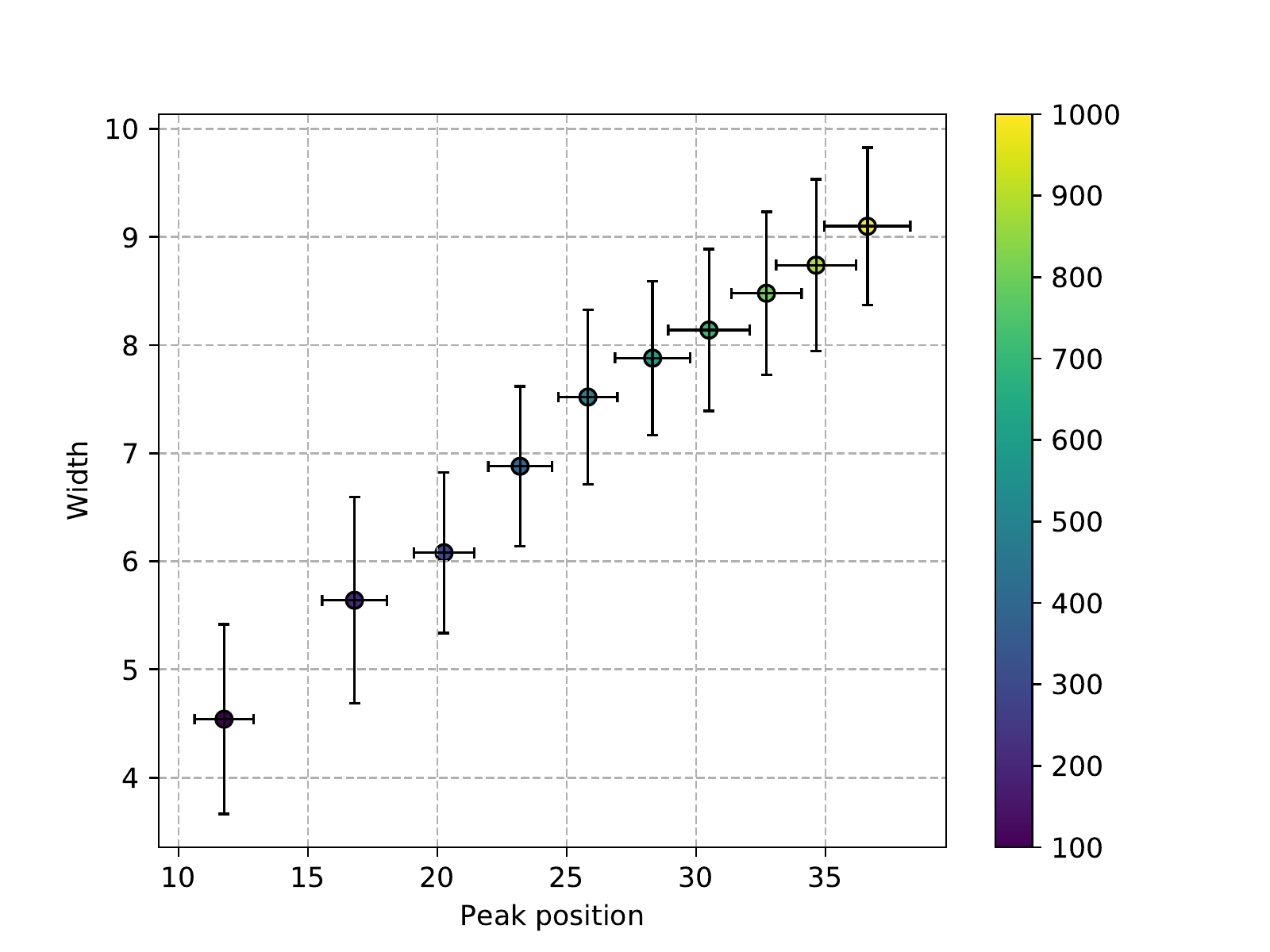}
\caption{A plot of width vs peak position for a variety of sizes of sprinkled causal sets.\\ The color indicates the number $N$ of elements in the causal set. \label{param}}
\end{figure}

\section{\sc Manifoldlike Causal Sets}

One of the motivations for this work was to explore the possibility of using the mean path length between two causal set elements $p$ and $q$ for a known value for the volume of $I(p,q)$ as a dimension estimator, similar to the use of the longest path length in Ref.\ \cite{meyer}, with the possible computational advantage that sampling the set of paths between $p$ and $q$ and using an average length to estimate the mean may be easier than finding the longest path. From simulations whose results are shown in this paper, as well as simulations in higher-dimensional Minkowski space, it appears that the average length of a sample of a few paths is indeed a valid dimension estimator, though it is unclear whether it is computationally better than the longest path method. One benefit of our approach and similar ones using the distribution of path lengths, however, is that it provides a criterion of manifoldikeness for causal sets.

\begin{figure}
\begin{minipage}{0.45 \textwidth}
\centering
\includegraphics[width = 6cm]{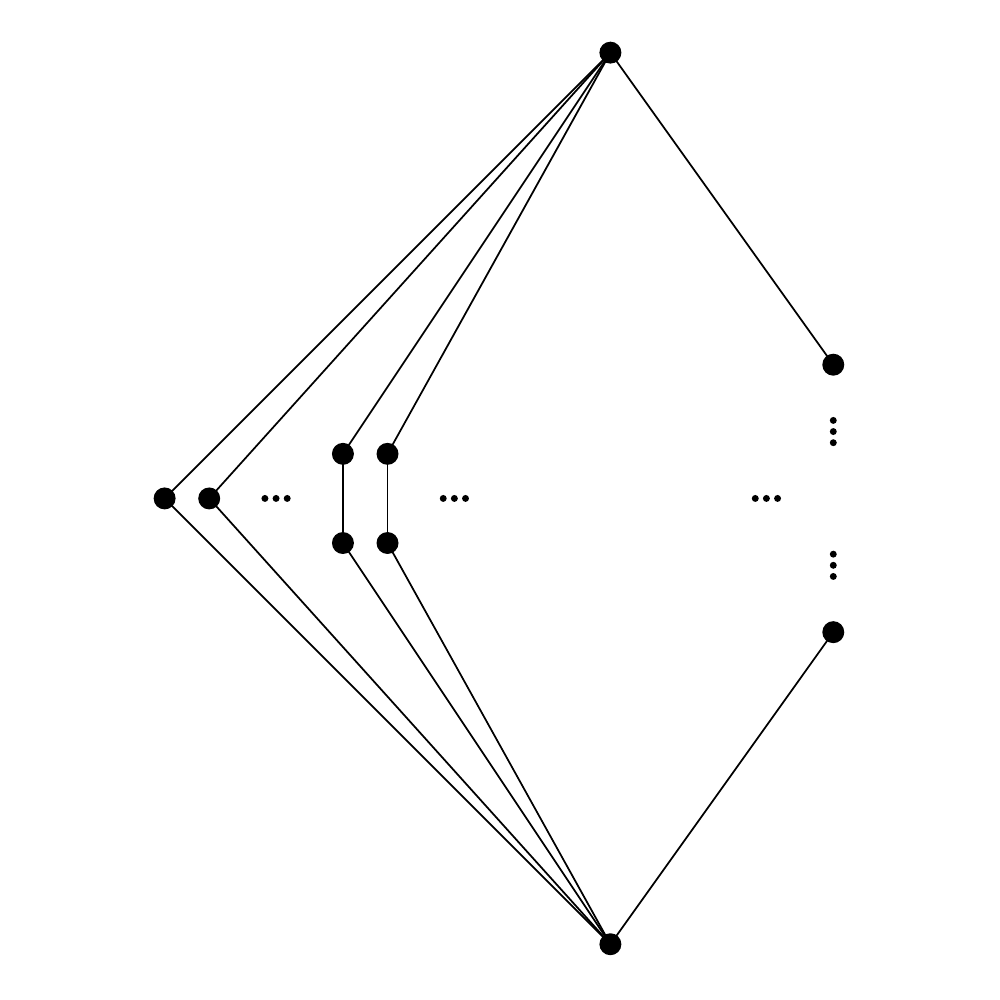}
\end{minipage}
\begin{minipage}{0.45 \textwidth}
\centering
\includegraphics[width = 6cm]{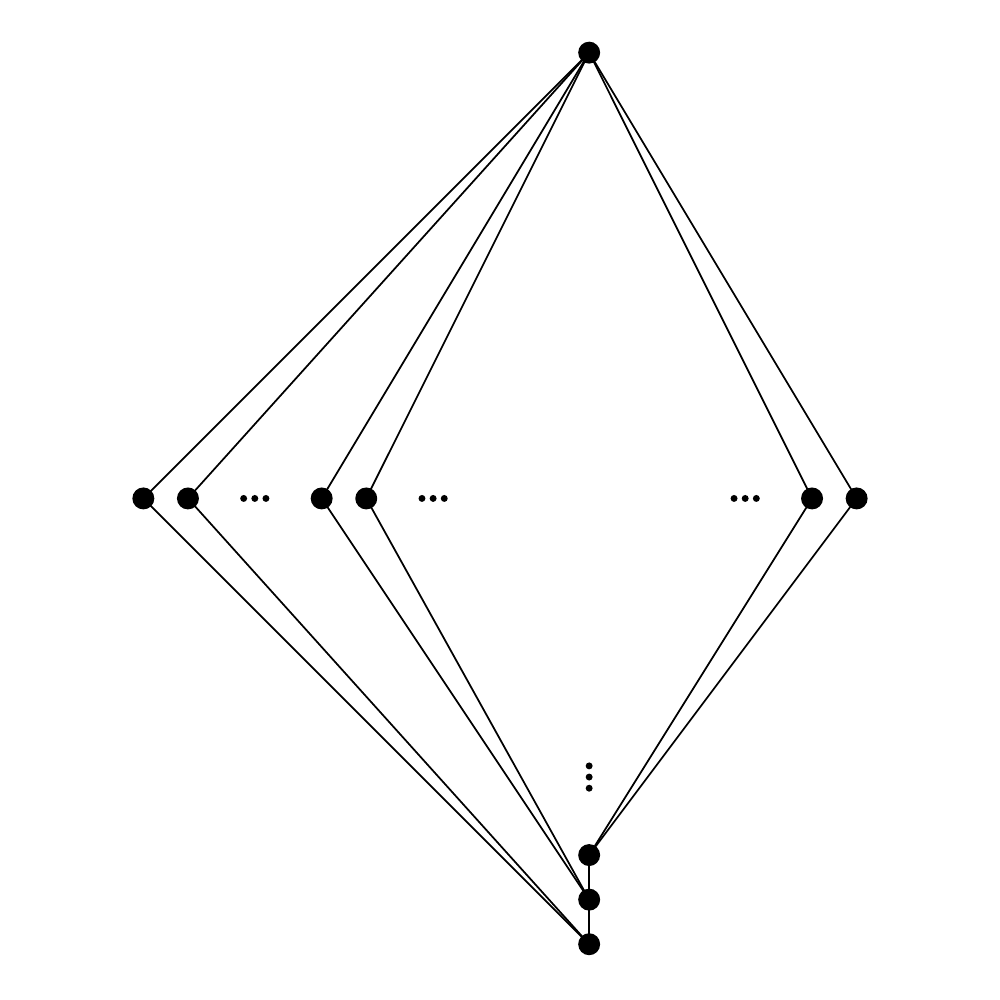}
\end{minipage}
\caption{Left: An artificially produced causal set whose purpose is to mimic both the height and width of our path length distribution. Right: A similarly produced artificial causal set which eliminates some redundancies to reduce the number of points while maintaining the path length distribution; however, it still has far more points than its manifoldlike cohorts. \label{art}}
\end{figure}

It is clear even from simple examples that quantities like the longest or the mean path length may be good dimension estimators only for causal sets known to be manifoldlike, and do not by themselves distinguish those causal sets from non-manifoldlike ones. For example, the union of $m$ chains of length $k$ with minimal points and maximal points identified is a causal set that can always be embedded in 2D Minkowski space, but adjusting the values of $m$ and $k$ one can obtain a relationship between the total number of elements $N = m(k-1)+2$ and the longest or mean path length $k$ that reproduces that of any Minkowski dimensionality. Similarly, a causal set could be constructed as the union of separate paths of various lengths all sharing the same minimal and maximal element, and with no other overlap, as in the left side of Fig.\ \ref{art}, with the number of chains of each length adjusted in a way that exactly matches the mean and width of the typical manifoldlike distribution. However, while the construction would yield the right value of $n_k$ for any length $k$, the total number of points in the causal set, $N=\sum_{k=2}^{k_{\text{max}}} n_k(k-1)+2,$ would be quite different as the manifoldlike distribution would have many paths sharing points and this contrived example does not. We could make the example slightly more realistic by forcing the paths to share all points not linked to the maximum point as in the right side of Fig.\ \ref{art}. This would limit the number of points significantly, with a total of $N = \sum_{k = 2}^{k_{\text{max}}} n_k+k_{\text{max}}$, where ${k_{\text{max}}}$ is the length of the longest path in the causal set. However, for $N\gg1$ this may still have several orders of magnitude more points than a manifoldlike causal set, as we can see by considering for example the right side of Fig.\ \ref{sprink}. If we use the average values of these 100-point sprinklings, the first method requires around $4\times 10^5$ points while the second one requires around $4\times 10^4$ points.

What we propose as a first manifoldlikeness criterion based on the distribution of path lengths $n_k$ is simply that any $N$-element causal set for which the mean value $k\_0$ and the width $\Delta$ of that distribution are not consistent with the corresponding theoretical values within statistical fluctuations cannot be manifoldlike. Based on the few examples we just saw, finding nonmanifoldlike causal sets that satisfy this condition is not trivial. Nevertheless, this condition is most likely not a sufficient one for manifoldlikeness. To further explore which causal sets meet or do not meet our criterion we will now provide two other types of examples of nonmanifoldlike causal sets.

One type includes causal sets that are not manifoldlike but are interesting for other reasons, and fail our criterion. The first example is a causal set that has one maximal and one minimal element, with all other elements located between them and unrelated to each other (i.e., one large antichain with added minimal and maximal elements); the path length distribution is a Kronecker delta $n\_k = \delta\_{k,2}$, with a sharp peak at length $2$ and zero for other lengths. A regular ``diamond lattice" (shown in the left side of Fig.\ \ref{example}) also has a path distribution sharply peaked at some length $k\_0 \approx \sqrt N$, with no paths of other lengths. More generally, most randomly chosen causal sets of $N$ elements with $N\gg1$ will look like the 3-layer Kleitman and Rothschild limit \cite{kleit} shown in the right side of Fig.\ \ref{example}, in which the first and third layers have $N/4$ elements, each of which is related to about half of the $N/2$ elements in the second layer; for such a causal set again $n\_k = \delta\_{k,2}$. The causal sets in these examples are all nonmanifoldlike, as one would not obtain them from uniform distributions of points in a Lorentzian manifold.

\begin{figure}
\begin{minipage}{0.45\textwidth}
\centering
\includegraphics[width=6cm]{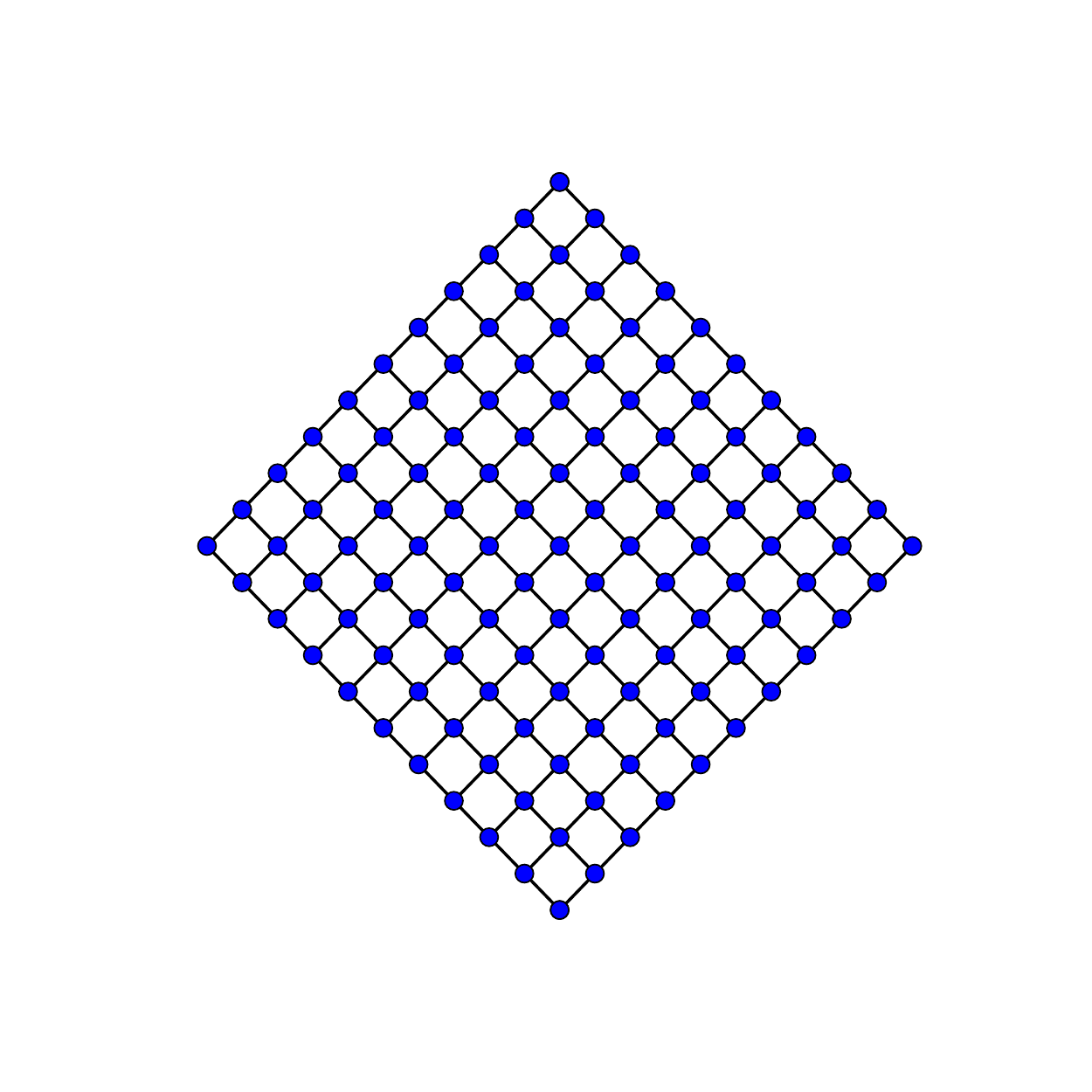}
\end{minipage}
\begin{minipage}{0.45 \textwidth}
\centering
\includegraphics[width=\textwidth]{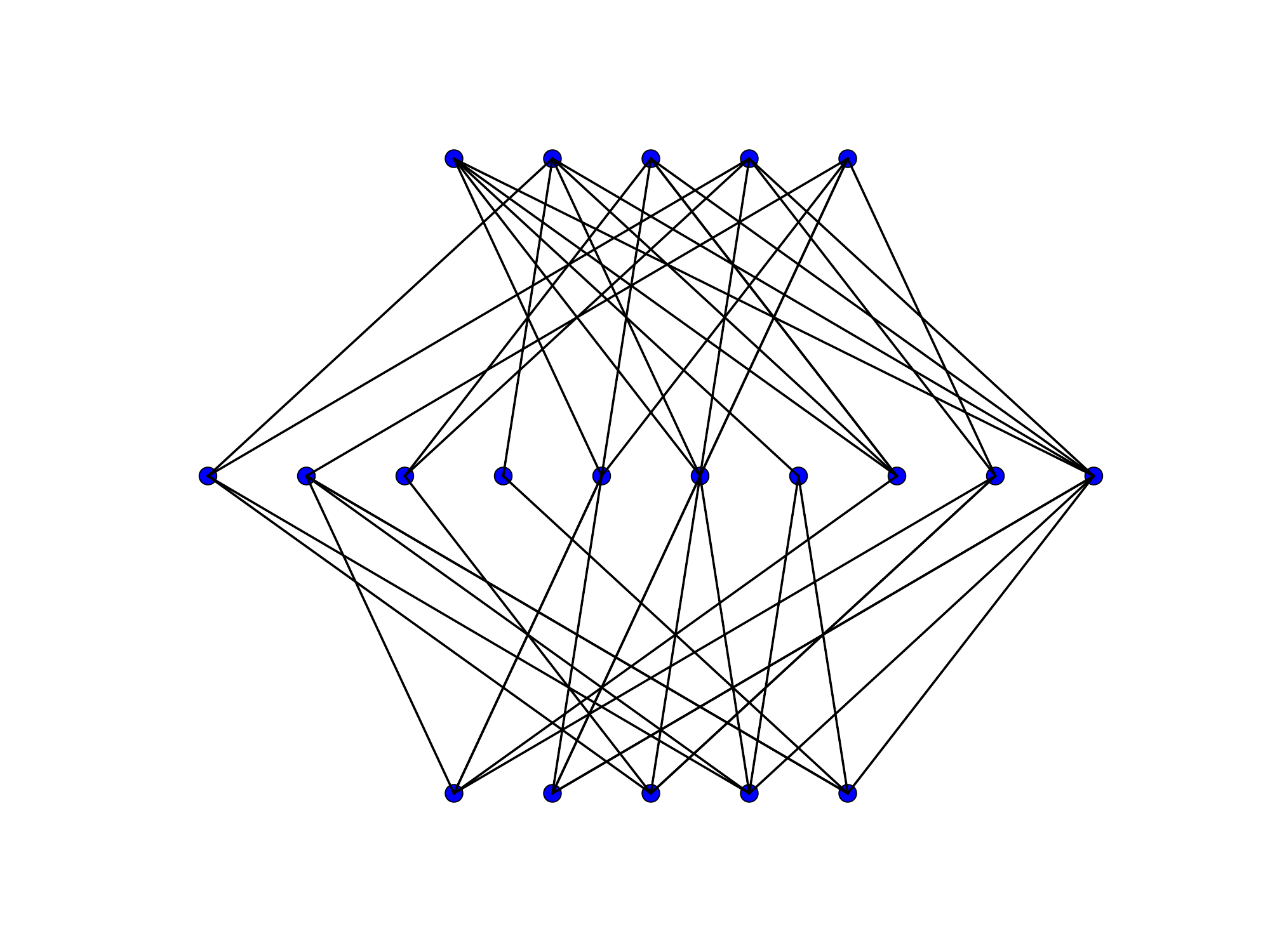}
\end{minipage}
\caption{Left: A regular lattice of $11^2 = 121$ points. The path distribution for this set is also sharply peaked, at $k = 20$ in this case, and not at all similar to that of a causal set from a random distribution of spacetime points. Right: A 20-element causal set illustrating the Kleitman-Rothschild limit, which has a sharp peak for path length $k = 2$.\label{example}}
\end{figure}

The other type of examples includes causal sets that are still not manifoldlike, but are likely to meet our criterion for manifoldlikeness. Fig.\ \ref{extrapoint} shows the effect of adding one extra point to a causal set obtained from a 250-point random sprinkling in an Alexandrov set $I(p,q)$ of 2D Minkowski space. The added point was to the future of a point approximately in the middle of the sprinkled causal set, and linked directly to the maximal element $q$; the left side of the figure shows the resulting augmented causal set. Because the added point gives rise to additional paths which are shorter than the ones that go through the original, sprinkled causal set, the new path length distribution will exhibit an additional small bump with a peak length shorter than the overall $k_0$. The right side of the figure shows the difference between the new path length distribution and the one without the additional point. This difference is very small compared to the overall distribution, but the feature it shows may be identifiable as characteristic of this particular type of nonmanifoldlike causal set.

We plan to continue studying the relationship between our criterion for manifoldlikeness and different ways in which a causal set may fail to be manifoldlike, to establish which additional criteria are needed to exclude nonmanifoldlike causal sets that are not physically interesting, and possibly formulate a more quantitative way to take into account statistical fluctuations in the path length distribution that would allow us to include causal sets which, strictly speaking, are not faithfully embeddable in a Lorentzian manifold, but may be physically interesting and we may want to call manifoldlike (see, e.g., \cite{henson}).

\begin{figure}
\begin{minipage}{0.45 \textwidth}
\centering
\includegraphics[width=6cm]{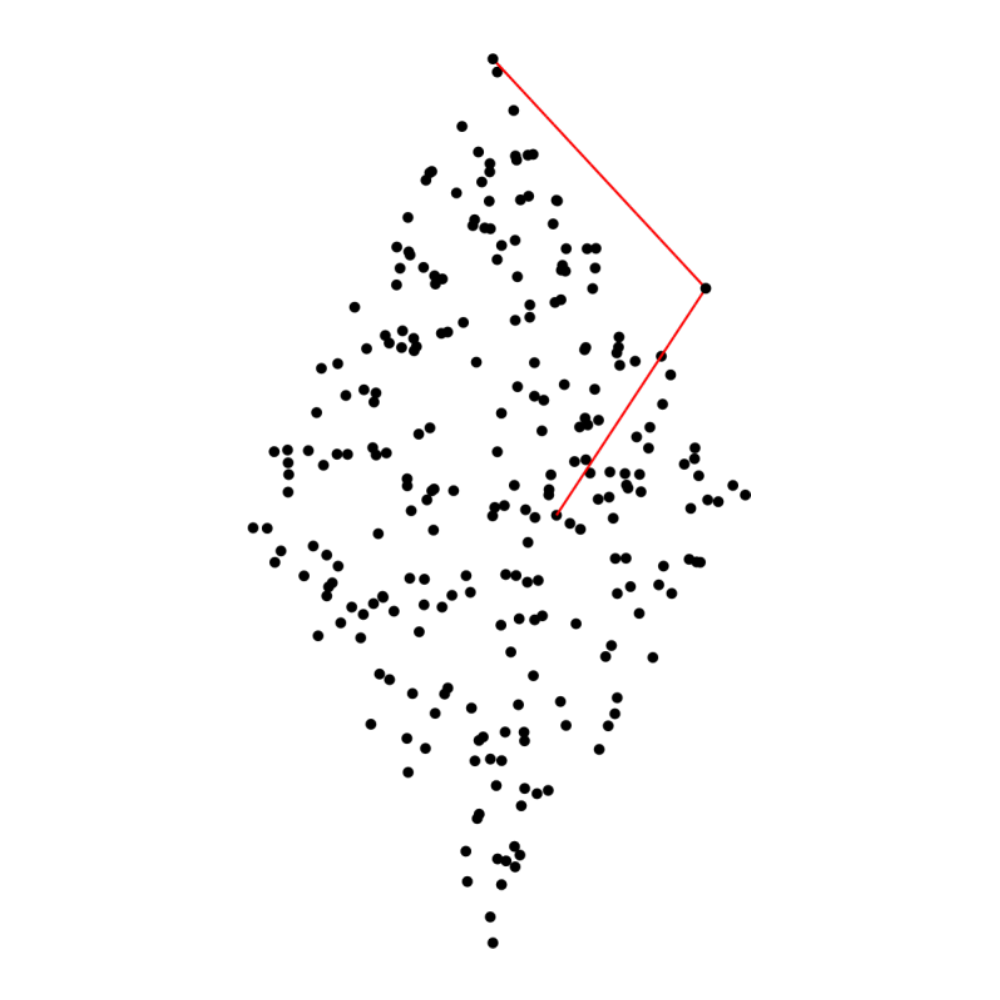}
\end{minipage}
\begin{minipage}{0.45\textwidth}
\centering
\includegraphics[width=\textwidth]{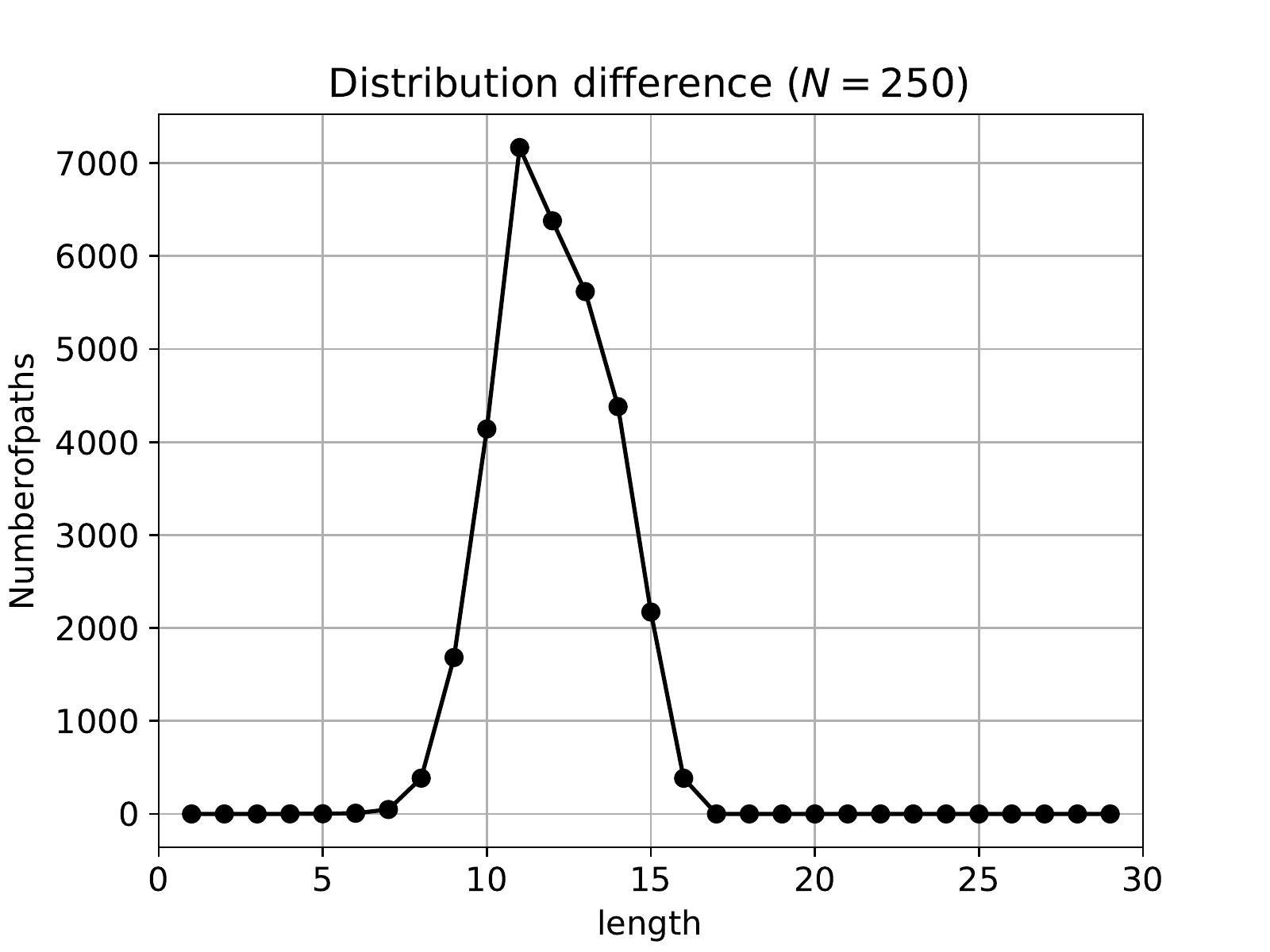}
\end{minipage}
\caption{Left: Causal set obtained from sprinkling in 2D Minkowski space and one added point with ``non-local" links. Right: Difference between the path length distributions.\label{extrapoint}}
\end{figure}

The results provided here will help us set up a procedure to establish whether a causal set is close to a spacetime manifold and address, at least in 2 dimensions, one of the most fundamental questions in causal set theory. The results for higher dimensions remain to be investigated in a future work and may have other applications, for instance in the expression for the Green's function for scalar fields propagating on causal sets in 4 dimensions \cite{Nomaan}. Last but not least, this work provides a relation between the most probable path length and the proper time in the continuum, which can be used to discretize the action written in Refs.\ \cite{sverd01, sverd02} and was not known in general.

\end{document}